\documentclass[12pt]{article}
\usepackage{amssymb}
\usepackage{amsmath}
\usepackage{apacite}
\usepackage{algorithm}
\usepackage{algpseudocode}
\usepackage{latexsym}
\usepackage{epsfig}
\usepackage{graphicx}
\usepackage{subfigure}
\usepackage{wasysym}
\usepackage{threeparttable}
\usepackage{natbib}
\usepackage{color}
\usepackage{epstopdf}
\usepackage{bm}
\usepackage{float}
\usepackage{todonotes}
\usepackage{verbatim}
\usepackage{svg}
\usepackage{amsmath}
\usepackage{booktabs}
\usepackage{placeins}
\usepackage{float}

\usepackage{hyperref}

\def\log{\hbox{log}}

\def\boxit#1{\vbox{\hrule\hbox{\vrule\kern6pt
          \vbox{\kern6pt#1\kern6pt}\kern6pt\vrule}\hrule}}

\def\bse{\begin{eqnarray*}}
\def\ese{\end{eqnarray*}}
\def\be{\begin{eqnarray}}
\def\ee{\end{eqnarray}}
\def\bq{\begin{equation}}
\def\eq{\end{equation}}
\def\bse{\begin{eqnarray*}}
\def\ese{\end{eqnarray*}}

\begin{document}

\thispagestyle{empty} 
\baselineskip=28pt

\begin{center}
{\LARGE{\bf A Bayesian Approach to Unit-level Dependent Multi-type Survey Data}}

\end{center}

\baselineskip=12pt

\vskip 2mm
\begin{center}
Zewei Kong\footnote{(\baselineskip=10pt to whom correspondence should be addressed) Department of Statistics and Data Science, University of Missouri, 146 Middlebush Hall, Columbia, MO 65211-6100, zk3bx@mail.missouri.edu},
Paul A. Parker\footnote{\baselineskip=10pt Department of Statistics, University of California Santa Cruz, 1156 High Street, Santa Cruz, CA 95064,
paulparker@ucsc.edu}, 
Jonathan R. Bradley \footnote{\baselineskip=10pt Department of Statistics and Data Science, University of Missouri,
134J Middlebush Hall, Columbia, MO 65211-6100, bradleyjr@missouri.edu},\,
    and  Scott H. Holan\footnote{\baselineskip=10pt Department of Statistics and Data Science, University of Missouri, 146 Middlebush Hall, Columbia, MO 65211-6100, holans@missouri.edu}
\\
\end{center}
%
%
%
%
\vskip 4mm
\baselineskip=12pt 
\begin{center}
{\bf Abstract}
\end{center}

The American Community Survey (ACS) Public Use Microdata Sample (PUMS) provides access to a wide range of unit-level survey data consisting of correlated Gaussian and binomial distributed survey responses along with associated survey weights. As such, we propose a Bayesian hierarchical framework for jointly modeling unit-level Gaussian and binomial survey data. The model introduces a shared area-level random effect to capture dependence across responses. Informative sampling is addressed using a pseudo-likelihood construction, and P\'{o}lya--Gamma data augmentation provides an efficient  conjugate Gibbs sampler, enabling scalable inference for large survey datasets. Through empirical simulations based on ACS PUMS data, we show that the joint model achieves notable reductions in mean squared error and improved interval scores compared to univariate and design-based estimators. Applying the method to the 2023 Illinois PUMS data, we find that the proposed multi-type model yields small-area estimates similar to those from the univariate model and the Horvitz--Thompson estimator, but with smaller posterior variances. The computational cost associated with the joint model is also comparable to that of the univariate binomial model. Combined with the empirical simulation results, these findings demonstrate the practical advantages of the proposed approach.

\baselineskip=12pt
\par\vfill\noindent
\par\medskip\noindent
\clearpage\pagebreak\newpage \pagenumbering{arabic}
\baselineskip=24pt

\section{Introduction}\label{sec:intro}

%

Policymakers and subject-matter experts often require precise summary statistics for small geographic areas or granular sub-populations. Federal surveys such as the American Community Survey (ACS) serve as an invaluable tool for this purpose. For example, the ACS helps to inform the Small Area Income and Poverty Estimates (SAIPE) program of the U.S. Census Bureau, which was developed to produce county- and school-district-level estimates of income and poverty {\citep{bell16}}. This program aims to provide annual income and poverty estimates for all counties and school districts, and these data inform allocations of billions of dollars in federal funds each year for key areas like education and healthcare. Similarly, under the U.S. Department of Housing and Urban Development (HUD), housing affordability assessments often rely on local poverty and income statistics. However, at these small spatial scales, direct design-based estimators are often not precise because the sample sizes are small.

To address the challenge of producing accurate statistics, the small area estimation (SAE) literature has increasingly incorporated unit-level models as an alternative to traditional area-level approaches. Compared to area-level models that rely only on aggregated summary statistics, unit-level models use unit-level microdata along with survey weights. These models have the capacity to capture within-area variability and allow covariate-outcome relationships to be modeled directly. This is particularly relevant for non-Gaussian response types such as binary outcomes (e.g., poverty status).
 For example, \cite{MolinaRao2010} developed a unit-level method for estimating poverty and income inequality measures, and \cite{hobza2018small} extended this framework to temporal binomial-logit mixed models, highlighting the importance of modeling binary outcomes such as poverty status. 

The goal of this paper is to introduce a new statistical method for Gaussian and binomial distributed unit-level survey data that accounts for informative sampling and capitalizes on cross-variable dependence to improve the precision of the statistics being estimated. Unit-level SAE must account for complex survey designs; when inclusion probabilities are related to the outcomes, ignoring the sampling mechanism can lead to biased inference. \cite{hidiroglou2016comparison} provided a design-based simulation comparison of unit-level and area-level estimators and showed that unit-level approaches can yield improved interval performance (e.g., coverage rate and interval length) under informative designs. \cite{ParkerJanickiHolan2023} gives an overview of approaches for accounting for informative sampling for unit-level models; however, one particularly useful approach in the Bayesian setting is the use of a weighted pseudo-likelihood or pseudo-posterior formulation (e.g., \cite{SavitskyToth2016}; see also \cite{parker2022computationally}  for an example using binary data). 

{One limitation of the current pseudo-posterior and related unit-level literature is that many developments remain within a single response family. For continuous outcomes, survey-weighted unit-level nested error linear regression models have been studied by \cite{you2002pseudo}. For count data, \cite{parker2020conjugate} developed a Bayesian unit-level framework under informative sampling. More recently, \cite{bugallo2024small} proposed unit-level multinomial mixed models, which likewise remain within a single distributional family.} As a result, there is still limited development of unit-level SAE frameworks that jointly model mixed-type outcomes, in particular Gaussian and binomial responses, while remaining computationally tractable and practically implementable.

\begin{figure}[htbp]
    \centering
    \includegraphics[width=1\linewidth]{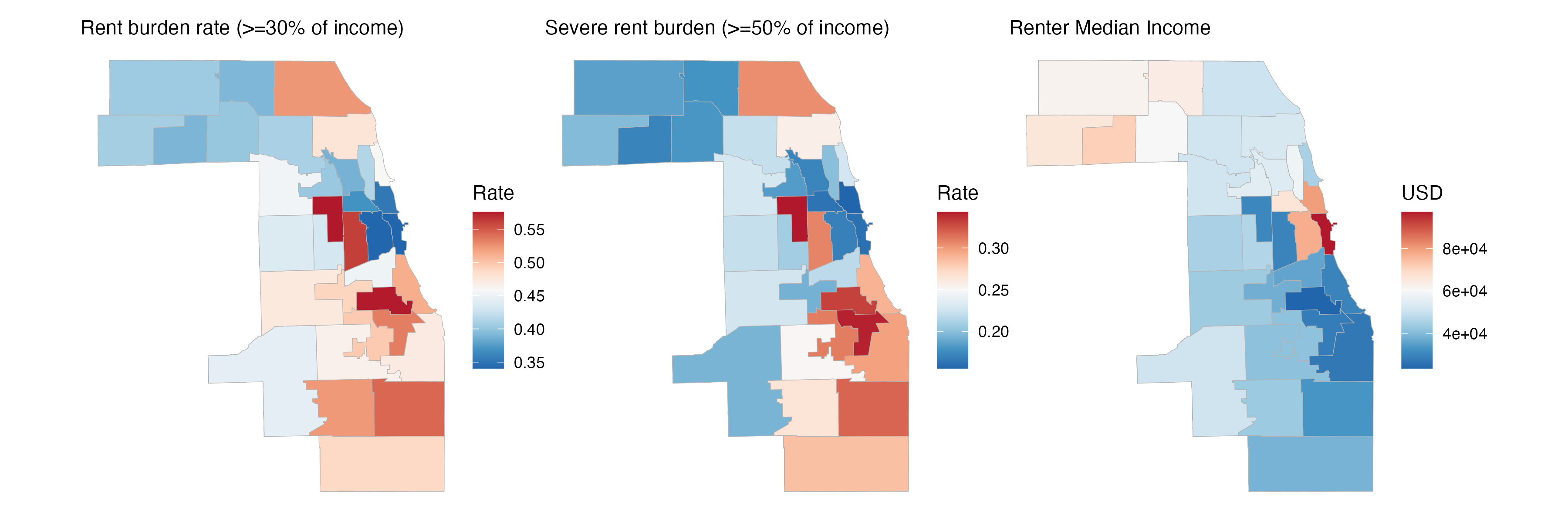}

    \vspace{0.8em}

    \includegraphics[width=0.9\linewidth]{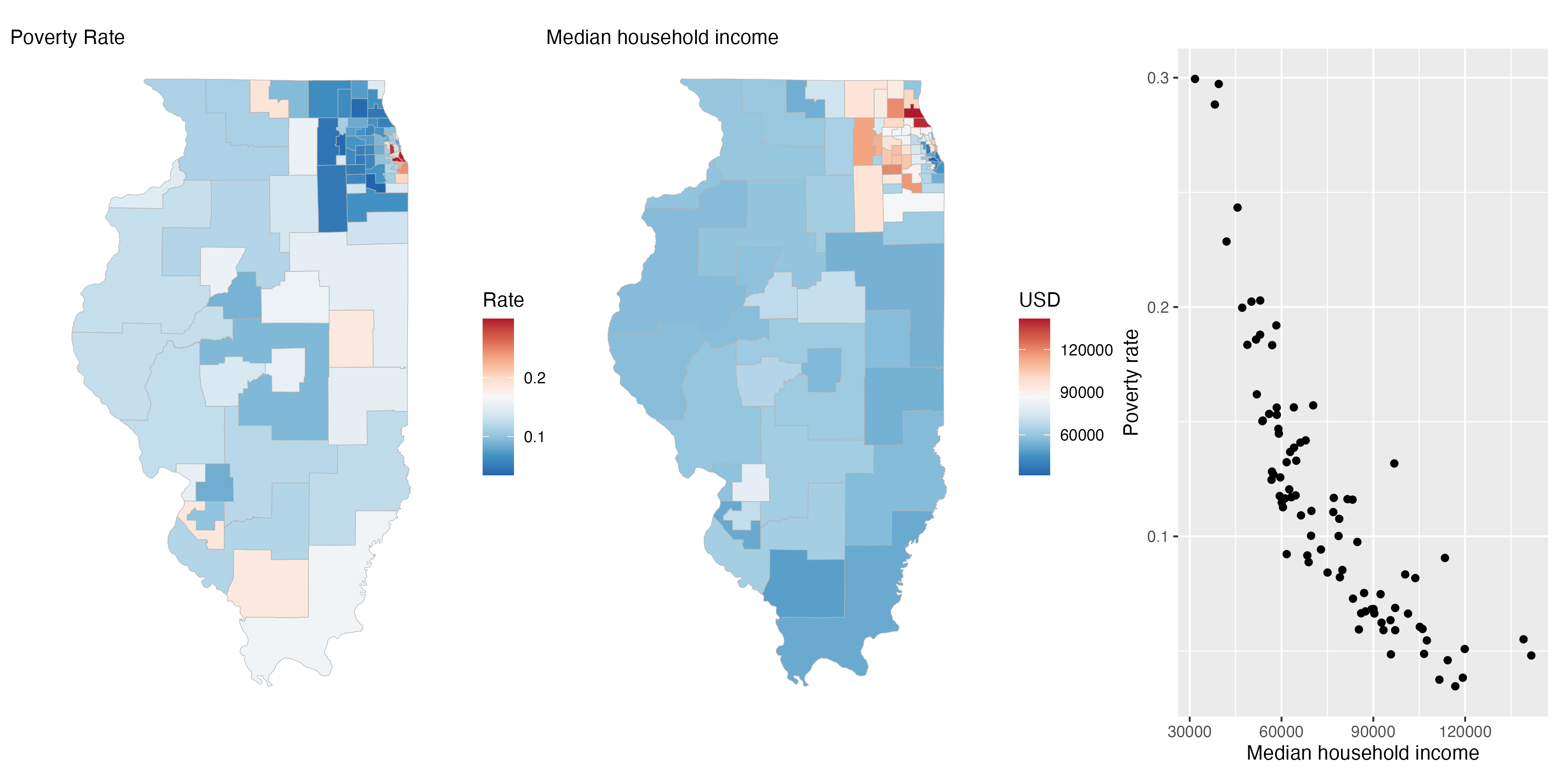}
    \caption{Spatial patterns of rent burden, poverty, and income across PUMAs. The top panel shows the spatial distribution of moderate rent burden (rent burden rate $\geq 30\%$ of income), severe rent burden (rent burden rate $\geq 50\%$ of income), and renter median income in the Chicago metropolitan area. The bottom panel shows the direct estimates of poverty rate and median household income at the Illinois PUMA level, together with the corresponding scatter plot between poverty rate and median household income.}
    \label{fig:spatial_combined}
\end{figure}

As a result, extending the pseudo-posterior formulation to the ``Multi-type'' framework is a key contribution of this work. By ``Multi-type'' we mean responses that are distributed from multiple types of distributions. Incorporating multivariate dependence in general is a strategy well-known to improve the precision of estimates. For instance, household income (or personal income) and poverty status have a strong natural correlation, as knowledge of one response innately informs the value of the other. Using the PUMA-level data from the Chicago metropolitan area, we can observe clear empirical patterns: areas with lower renter income tend to have higher rent-burden rates (Figures~\ref{fig:spatial_combined}), and areas with lower overall income tend to have higher poverty rates. These patterns also exhibit geographic heterogeneity across areas.

While it is straightforward to see that incorporating such multivariate dependence across data types can improve the precision of estimates it is less straightforward to develop an efficient strategy to model such data. In general, joint modeling of multi-type data is an extremely active modern area of study, as many are recognizing the limitations of ``unit-type'' modeling for non-survey data. Several algorithms make use of computationally inefficient Markov chain Monte Carlo (MCMC) algorithms that require Metropolis-Hastings steps \citep[see][among others]{christensen2002,gs2,wu2015bayesian,leClarke,Todd} for non-survey data. Other strategies make use of regularized loss functions \citep{allen,ekvall2022mixed}, multi-task learning models \citep{Argyriou,Kim,Yang}, copulas \citep{rank2,rank1, copula1,copula2}, and tree based methodology \citep{forest} to jointly model multi-type data. However, these strategies do not address uncertainty via a Bayesian perspective, and do not account for informative sampling. There are Bayesian strategies that make use of conjugate distributions to model multi-type data from the exponential family that either avoid Metropolis-Hastings steps \citep{bradley2022joint,nandy2022bayesian} or MCMC and approximating the posterior distribution \citep{clinch2026exact}. However, these approaches do not incorporate informative sampling, and it is less clear how this can be done in the pseudo-posterior framework due to the presence of an additive discrepancy term. To address the computational challenges of jointly modeling multiple data types we consider a conjugate strategy that incorporates P\'{o}lya–Gamma data augmentation \citep{polson2013bayesian} to obtain conditionally conjugate updates for the binomial component. This leads to an efficient Gibbs sampling strategy that does not require Metropolis-Hastings steps.


{A second challenge is that, although multivariate spatial frameworks have been developed for survey and official statistics, they are often formulated and implemented for area-level or other aggregated responses. For example, \citet{porter2015small} developed multivariate spatial Fay--Herriot models with latent spatial dependence, and \citet{bradley2015multivariate} and \citet{bradley2016multivariate} developed multivariate spatio-temporal frameworks for high-dimensional areal data and survey fusion, respectively. However, these contributions primarily operate on area-level or aggregated survey responses, and extending such ideas to unit-level settings with mixed outcome types is nontrivial. In our setting, one must account simultaneously for cross-outcome dependence, area-level dependence, mixed data types, and informative sampling at the unit level. Although \citet{SavitskyToth2016} provide a Bayesian justification for pseudo-likelihood weighting under informative sampling, incorporating this idea into a mixed-type multivariate unit-level framework remains methodologically and computationally challenging.}

Our strategy is to develop a novel Bayesian hierarchical framework for jointly modeling unit-level Gaussian outcomes (income) and binomial outcomes (poverty status). Our approach has three key components. First, we introduce a shared area-level random effect to induce dependence across {data types and areas}. Second, we incorporate informative sampling through a survey-weighted pseudo-likelihood using scaled weights, enabling Bayesian inference that accounts for unequal selection probabilities without explicitly modeling the sampling design. Third, for the binomial component, we use the P\'{o}lya–Gamma data augmentation of \cite{polson2013bayesian} to obtain conditionally conjugate updates and implement an efficient Gibbs sampling algorithm.

The remainder of this paper is organized as follows. Section~\ref{sec:method} introduces the proposed joint Gaussian–binomial unit-level model, detailing the pseudo-likelihood construction and the Gibbs sampler, while contrasting the {proposed} framework with separate univariate baselines. Section~\ref{sec: simulation} presents an empirical simulation study that evaluates the performance of point and interval estimators under varying levels of sample sparsity and design informativeness. Section~\ref{sec:data} applies the method to ACS public-use microdata, demonstrating practical gains in predictive precision and reductions in posterior variance. Finally, Section~\ref{sec:conclusion} provides a discussion.

\section{Multi-Type Unit-Level Model}\label{sec:method}
Let $S$ denote the set of sampled units. For unit $i$, let $\pi_i=\Pr(i\in S)$ denote the inclusion probability and let $w_i$ denote the corresponding inverse-probability design weight. To account for unequal selection probabilities without explicitly modeling the sampling mechanism, we incorporate these weights through a pseudo-likelihood framework. Posterior inference is conducted via MCMC. For the binomial response, we use P\'{o}lya--Gamma data augmentation to obtain Gibbs updates.

\subsection{Pseudo‐likelihood Under Informative Sampling}

In complex surveys such as the ACS, sampling probabilities vary across units and can depend on features of the population such as income, education level, or place of residence. When $\pi_i$ depends on the response, the sample is informative and the ordinary likelihood 
\[
L(\theta;Z)=\prod_{i\in S}f(Z_i\mid\theta)
\]
can yield biased inference.

To prevent posterior over-concentration caused by raw weights, we rescale the survey weights \citep{SavitskyToth2016}. This rescaling preserves relative weights while maintaining the total information on the order of the sample size. The scaled weights are defined as
\[
\tilde w_i \;=\;\frac{n\,w_i}{\sum_{j\in S}w_j},
\]
so that $\sum_{i\in S}\tilde w_i=n$.  The pseudo-likelihood is then
\[
L_{\mathrm{PL}}(\theta;Z,w)\;=\;\prod_{i\in S}f(Z_i\mid\theta)^{\tilde w_i},
\]
and the corresponding pseudo‐posterior is
\[
\pi_{\mathrm{PL}}(\theta\mid Z,w)\;\propto\;L_{\mathrm{PL}}(\theta;Z,w)\,\pi(\theta).
\]

\subsection{Model Specification}

We first establish univariate pseudo-likelihood models for Gaussian and binomial responses. These univariate models serve as baseline comparators for the joint multi-type model. They allow us to evaluate whether incorporating cross-response dependence improves estimation performance.

\subsubsection{Univariate Gaussian response}
\label{sec:uni_gaussian}

Let $\{(x_i,\boldsymbol{\phi}_i,Z_{1,i},w_i)\}_{i=1}^n$ denote a sample of $n$ units from $r$ areas. For the Gaussian response $Z_{1,i} \in \mathbb{R}$, let $x_i \in \mathbb{R}^p$ be the covariate vector with coefficients $\beta_1$,  let $\boldsymbol{\phi}_i$ be the corresponding area-level design vector and let $D=\mathrm{Diag}(\tilde w_1/\sigma^2,\ldots,\tilde w_n/\sigma^2)$.

The Gaussian pseudo-likelihood model is
\[
Z_{1,i} \mid \beta_1, U, \sigma^2 \sim N(\mu_{1,i},\,\sigma^2), 
\,\,\,\, 
\mu_{1,i} = x_i^\top \beta_1 + \boldsymbol{\phi}_i^\top U,
\]
where $U \in \mathbb{R}^q$ is the area-level random effect. We assign the following priors:
\[
\beta_1 \sim N(0,\sigma_\beta^2 I_p), \,\,\,\,
U \sim N(0,\sigma_u^2 I_q), \,\,\,\,
\sigma^2 \sim \mathrm{InvGamma}(a_\varepsilon,b_\varepsilon), \,\,\,\,
\sigma_u^2 \sim \mathrm{InvGamma}(a_u,b_u).
\]
Let $X = [x_1^\top;\ldots;x_n^\top]$, 
$\Phi = [\boldsymbol{\phi}_1^\top;\ldots;\boldsymbol{\phi}_n^\top]$, 
and $Z_1 = (Z_{1,1},\ldots,Z_{1,n})^\top$.
Under the pseudo-likelihood, full conditional distributions are conjugate and given by:
\begin{align*}
\beta_1 \mid \cdot &\sim 
N\!\Bigl(
V_{\beta_1} X^\top D (Z_1 - \Phi U),\;
V_{\beta_1}
\Bigr), \quad
V_{\beta_1} = \bigl(X^\top D X + \sigma_\beta^{-2} I_p\bigr)^{-1},\\[2mm]
U \mid \cdot &\sim 
N\!\Bigl(
V_U \Phi^\top D (Z_1 - X \beta_1),\;
V_U
\Bigr), \quad
V_U = \bigl(\Phi^\top D \Phi + \sigma_u^{-2} I_q\bigr)^{-1},\\[2mm]
\sigma^2 \mid \cdot &\sim 
\mathrm{InvGamma}\!\left(
a_\varepsilon + \tfrac{1}{2}\sum_{i=1}^n \tilde w_i,\;
b_\varepsilon + \tfrac{1}{2}\sum_{i=1}^n \tilde w_i (Z_{1,i}-\mu_{1,i})^2
\right),\\[2mm]
\sigma_u^2 \mid \cdot &\sim
\mathrm{InvGamma}\!\left(
a_u + \tfrac{q}{2},\;
b_u + \tfrac{1}{2}U^\top U
\right).
\end{align*}

\subsubsection{Univariate Binomial response with P\'{o}lya--Gamma augmentation}
\label{sec:uni_binomial}
For the binomial response model, we use the P\'{o}lya--Gamma (PG) data augmentation approach of \citet{polson2013bayesian}. Conditional on latent PG variables, the logistic likelihood can be written as a Gaussian kernel, which yields conjugate full conditional distributions under the pseudo-likelihood.
If $\omega \sim \mathrm{PG}(b,0)$, then for any $\psi \in \mathbb{R}$ and $a \in \mathbb{R}$,
\[
\frac{e^{a\psi}}{(1+e^{\psi})^{b}}
= 2^{-b} e^{\kappa \psi}
\,\mathbb{E}_{\omega}\!\left[
\exp\!\left(-\tfrac12 \omega \psi^{2}\right)
\right],
\,\,\,\,
\kappa = a - \tfrac{b}{2},
\]
and the conditional distribution $\omega \mid \psi \sim \mathrm{PG}(b,\psi)$ is given in \citet{polson2013bayesian}.

Consider a binomial observation $y \sim \mathrm{Bin}(n,p)$ with $\mathrm{logit}(p) = \psi$. The likelihood can be written as
\[
\Bigl\{\tfrac{e^{\psi}}{1+e^{\psi}}\Bigr\}^{y}
\Bigl\{\tfrac{1}{1+e^{\psi}}\Bigr\}^{n-y}
=
\frac{e^{y\psi}}{(1+e^{\psi})^{n}},
\]
Under P\'{o}lya–Gamma augmentation, this likelihood can be written in a conditionally Gaussian form. For unit $i$, let $Z_{2,i} \in \{0,1\}$, or more generally $Z_{2,i} \sim \mathrm{Binomial}(n_i,p_i)$, with
\[
\mathrm{logit}(p_i) = \psi_i = x_i^\top \beta_2 + \boldsymbol{\phi}_i^\top U.
\]
Under the pseudo-likelihood, each binomial likelihood contribution is raised to the scaled survey weight $\tilde w_i = n w_i / \sum_{j=1}^n w_j$. Define
\[
b_i = \tilde w_i n_i,
\,\,\,\,
\kappa_i = \tilde w_i (Z_{2,i} - n_i/2).
\]
Conditional on latent variables $\omega_i$, the augmented likelihood takes the form
\[
\exp\!\{\kappa_i \psi_i\}
\exp\!\{-\tfrac12 \omega_i \psi_i^2\}
\;\propto\;
\exp\!\Bigl\{-\tfrac12 \omega_i (\psi_i - z_i)^2\Bigr\},
\,\,\,\,
z_i = \kappa_i / \omega_i,
\]
with $\omega_i \mid \psi_i \sim \mathrm{PG}(b_i,\psi_i)$. Let
$\Omega = \mathrm{Diag}(\omega_1,\ldots,\omega_n)$, 
$z = (z_1,\ldots,z_n)^\top$,
$X = [x_1^\top;\ldots;x_n^\top]$, 
and $\Phi = [\boldsymbol{\phi}_1^\top;\ldots;\boldsymbol{\phi}_n^\top]$. With priors $\beta_2 \sim N(0,\sigma_\beta^2 I_p)$, $U \sim N(0,\sigma_u^2 I_q)$, and $\sigma_u^2 \sim \mathrm{InvGamma}(a_u,b_u)$, the full conditional distributions are
\begin{align*}
\beta_2 \mid \cdot &\sim
N\!\Bigl(
V_{\beta_2} X^\top \Omega (z - \Phi U),\;
V_{\beta_2}
\Bigr), \quad
V_{\beta_2} = (X^\top \Omega X + \sigma_\beta^{-2} I_p)^{-1},\\[1mm]
U \mid \cdot &\sim
N\!\Bigl(
V_{U,2} \Phi^\top \Omega (z - X \beta_2),\;
V_{U,2}
\Bigr), \quad
V_{U,2} = (\Phi^\top \Omega \Phi + \sigma_u^{-2} I_q)^{-1},\\[1mm]
\omega_i \mid \cdot &\sim \mathrm{PG}(b_i,\psi_i), \qquad i = 1,\ldots,n,\\[1mm]
\sigma_u^2 \mid \cdot &\sim
\mathrm{InvGamma}\!\left(
a_u + \tfrac{q}{2},\;
b_u + \tfrac{1}{2}U^\top U
\right).
\end{align*}
In the Bernoulli case, $n_i \equiv 1$, so $b_i = \tilde w_i$ and $\kappa_i = \tilde w_i (Z_{2,i} - 1/2)$.

\subsubsection{Multi-type Unit-Level Model}

We retain the notation introduced above. For each sampled unit $i \in S$, 
$Z_{1,i} \in \mathbb{R}$ denotes the Gaussian response and 
$Z_{2,i} \in \{0,1\}$ (or more generally $Z_{2,i} \sim \mathrm{Binomial}(n_i,p_i)$) denotes the binomial response. The covariate vectors are $x_{1,i} \in \mathbb{R}^{p_1}$ and $x_{2,i} \in \mathbb{R}^{p_2}$. We also define 
$X_1=[x_{1,1}^\top;\ldots;x_{1,n}^\top]$, 
$X_2=[x_{2,1}^\top;\ldots;x_{2,n}^\top]$, and 
$\Phi=[\boldsymbol{\phi}_1^\top;\ldots;\boldsymbol{\phi}_n^\top]$.

The joint model includes a shared area-level random effect and a binomial-specific area-level random effect given by
\begin{eqnarray*}
Z_{1,i} &=& x_{1,i}^\top \beta_1 + \tau_1 \boldsymbol{\phi}_i^\top \eta + \varepsilon_i,
\,\,\,\,
\varepsilon_i \sim N(0,\sigma^2),\\
\mathrm{logit}(p_i) &=& x_{2,i}^\top \beta_2 + \boldsymbol{\phi}_i^\top \eta + \boldsymbol{\phi}_i^\top \zeta,
\,\,\,\,
Z_{2,i} \mid p_i \sim \mathrm{Binomial}(n_i,p_i).
\end{eqnarray*}
Table~\ref{tab:symbols} summarizes the main notation and meanings.
\begin{table}[ht!]
\centering
\caption{Symbols in the joint model and their interpretation.}
\label{tab:symbols}
\begin{tabular}{ll}
\toprule
Symbol & Meaning \\
\midrule
$Z_{1,i}$, $Z_{2,i}$ & Gaussian and binomial responses for unit $i$ \\
$x_{1,i}$, $x_{2,i}$ & Covariate vectors for the two outcomes \\
$\phi_i$ & area-level design vector for unit $i$ \\
$\beta_1$, $\beta_2$ & Fixed effects for Gaussian and binomial means \\
$\eta$ & Shared area-level effect; induces cross–response dependence \\
$\zeta$ & Binomial–specific area-level effect;  captures residual variation beyond $\eta$\\
$\tau_1$ & Scale on $\eta$ in the Gaussian block; controls borrowing from $Z_2$ to $Z_1$ \\
$\sigma^2$ & Gaussian residual variance \\
$\sigma_\eta^2$, $\sigma_\zeta^2$ & Variances of $\eta$ and $\zeta$ \\
$\Phi$ & Matrix stacking $\phi_i^\top$; maps area-level effects to units \\
$\tilde w_i$ & Scaled survey weight for unit $i$ used in the pseudo–likelihood \\
$W$ & Diagonal matrix of scaled survey weights, $\mathrm{Diag}(\tilde w_1,\ldots,\tilde w_n)$ \\
$D=W/\sigma^2$ & Weighted Gaussian precision matrix \\
$Q_G=\Phi^\top (W/\sigma^2) \Phi$ & Weighted Gaussian information for area-level effects \\
$Q_B=\Phi^\top \Omega \Phi$ & Weighted binomial (PG) information for area-level effects \\
\bottomrule
\end{tabular}
\end{table}
We assign Gaussian priors to the regression coefficients and area-level random effects:
\[
\beta_1 \sim N(0,\sigma_\beta^2 I_{p_1}), \,\,\,\,
\beta_2 \sim N(0,\sigma_\beta^2 I_{p_2}),
\]
\[
\eta \sim N(0,\sigma_\eta^2 I_q), \,\,\,\,
\zeta \sim N(0,\sigma_\zeta^2 I_q), \,\,\,\,
\tau_1 \sim N(0,\sigma_\tau^2).
\]
For the variance components, we use inverse-gamma priors
\[
\sigma^2 \sim \mathrm{InvGamma}(a_\varepsilon,b_\varepsilon), \,\,\,\,
\sigma_\eta^2 \sim \mathrm{InvGamma}(a_\eta,b_\eta), \,\,\,\,
\sigma_\zeta^2 \sim \mathrm{InvGamma}(a_\zeta,b_\zeta).
\]
In our implementation, we set $\sigma_\beta=1000$ and treat $\sigma_\tau^2$ as fixed. We use weakly informative inverse-gamma priors for $\sigma^2$, $\sigma_\eta^2$, and $\sigma_\zeta^2$. With the priors given above, the full conditional distributions are
\begin{align*}
\beta_1 \mid \cdot &\sim
N\!\Bigl(
V_{\beta_1}\,X_1^\top D (Z_1-\tau_1\Phi\eta),\;
V_{\beta_1}
\Bigr),
\qquad
V_{\beta_1}
=
\bigl(X_1^\top D X_1+\sigma_\beta^{-2}I_{p_1}\bigr)^{-1},
\\[1mm]
\beta_2 \mid \cdot &\sim
N\!\Bigl(
V_{\beta_2}\,X_2^\top \Omega (z-\Phi\eta-\Phi\zeta),\;
V_{\beta_2}
\Bigr),
\qquad
V_{\beta_2}
=
\bigl(X_2^\top \Omega X_2+\sigma_\beta^{-2}I_{p_2}\bigr)^{-1},
\\[1mm]
\eta \mid \cdot &\sim
N\!\Bigl(
V_{\eta}\Bigl[
\tau_1 \Phi^\top D (Z_1-X_1\beta_1)
+
\Phi^\top \Omega (z-X_2\beta_2-\Phi\zeta)
\Bigr],\;
V_{\eta}
\Bigr),
\\
&\qquad
V_{\eta}
=
\bigl(
\tau_1^2 Q_G+Q_B+\sigma_\eta^{-2}I_q
\bigr)^{-1},
\\[1mm]
\zeta \mid \cdot &\sim
N\!\Bigl(
V_{\zeta}\,\Phi^\top \Omega (z-X_2\beta_2-\Phi\eta),\;
V_{\zeta}
\Bigr),
\qquad
V_{\zeta}
=
\bigl(
Q_B+\sigma_\zeta^{-2}I_q
\bigr)^{-1},
\\[1mm]
\tau_1 \mid \cdot &\sim
N\!\Bigl(
V_{\tau_1}\,(\Phi\eta)^\top D (Z_1-X_1\beta_1),\;
V_{\tau_1}
\Bigr),
\qquad
V_{\tau_1}
=
\bigl(
(\Phi\eta)^\top D (\Phi\eta)+\sigma_\tau^{-2}
\bigr)^{-1},
\\[1mm]
\sigma^2 \mid \cdot &\sim
\mathrm{InvGamma}\!\left(
a_\varepsilon+\frac{1}{2}\sum_{i=1}^n \tilde w_i,\;
b_\varepsilon+\frac{1}{2}\sum_{i=1}^n \tilde w_i
\bigl(Z_{1,i}-x_{1,i}^\top\beta_1-\tau_1\phi_i^\top\eta\bigr)^2
\right),
\\[1mm]
\sigma_\eta^2 \mid \cdot &\sim
\mathrm{InvGamma}\!\left(
a_\eta+\frac{q}{2},\;
b_\eta+\frac{1}{2}\eta^\top\eta
\right),
\\[1mm]
\sigma_\zeta^2 \mid \cdot &\sim
\mathrm{InvGamma}\!\left(
a_\zeta+\frac{q}{2},\;
b_\zeta+\frac{1}{2}\zeta^\top\zeta
\right),
\\[1mm]
\omega_i \mid \cdot &\sim \mathrm{PG}(b_i,\psi_i),
\qquad i=1,\ldots,n,
\end{align*}
where
\[
\psi_i=x_{2,i}^\top\beta_2+\phi_i^\top\eta+\phi_i^\top\zeta.
\]
In the Bernoulli case, $n_i \equiv 1$, so
\[
b_i=\tilde w_i,\qquad
\kappa_i=\tilde w_i\left(Z_{2,i}-\frac{1}{2}\right).
\]
The shared random effect links the Gaussian and binomial components and allows information to be borrowed across the two responses. The binomial-specific random effect absorbs residual area-level variation not explained by the shared effect. Details for deriving the full conditional distribution of the shared random effect are given in Appendix.

\subsection{Adjacency-based basis specification}

The model above is written in terms of a general area-level design vector $\phi_i$. One useful choice is to construct $\phi_i$ from a lower-dimensional set of spatial basis functions, see \cite{parker2022computationally}. This provides a lower-dimensional representation of the spatial effect, captures spatial dependence across neighboring areas, and can reduce computational cost when the number of areas is large.

Let $A$ denote the $r \times r$ adjacency matrix of the areas, where $A_{ij}=1$ if areas $i$ and $j$ share a common boundary and $A_{ij}=0$ otherwise. Let $B$ be an $r \times q$ matrix of spatial basis functions constructed from the adjacency structure, with $q < r$. In our implementation, the columns of $B$ are given by the eigenvectors of the adjacency matrix associated with its positive eigenvalues.  An area-level indicator specification can be considered as a special case of this design-vector; see \cite{parker2020conjugate} for an example. For a sampled unit in area $a(i)$, we define
\[
\phi_i = B_{a(i), \cdot}^{\top}.
\]

Under this specification, the shared and outcome-specific spatial effects are represented in the lower-dimensional basis space. The Gaussian and binomial components become
\begin{eqnarray*}
Z_{1,i} &=& x_{1,i}^{\top}\beta_1 + \tau_1 \phi_i^{\top}\eta + \epsilon_i, \qquad \epsilon_i \sim N(0,\sigma^2),\\
\text{logit}(p_i) &=& x_{2,i}^{\top}\beta_2 + \phi_i^{\top}\eta + \phi_i^{\top}\zeta,
\qquad Z_{2,i}\mid p_i \sim \text{Binomial}(n_i,p_i),
\end{eqnarray*}
where $\eta \in \mathbb{R}^q$ and $\zeta \in \mathbb{R}^q$ are basis coefficients.

\subsection{Poststratification to Domains}
\label{sec:post}
Poststratification aggregates unit-level model outputs into area-level estimates using draws from the model’s posterior predictive distribution. Each area $k$ is divided into $J$ poststratification cells indexed by $j=1,\dots,J$. Each cell is defined by a unique combination of categorical covariates such as sex and education. For each cell $j$ in area $k$, let $N_{kj}$ denote the population size in that cell. Let $\mathcal{J}_k$ be the set of cells in area $k$, and let $N_k=\sum_{j\in\mathcal{J}_k} N_{kj}$ be the total population of area $k$. 

At each MCMC iteration $t$, we compute the expected outcome for each cell $j$ and aggregate these quantities using the normalized population weights $q_{kj}=N_{kj}/N_k$ for $j\in\mathcal{J}_k$. Repeating this over all iterations yields the area-level posterior distributions.

For the Gaussian outcome, the conditional cell mean at iteration $t$ is
\[
\theta_{1j}^{(t)}
= x_{1j}^\top \beta_1^{(t)} + \tau_1^{(t)} \phi_j^\top \eta^{(t)},
\]
where $x_{1j}$ is the Gaussian covariate vector for cell $j$, $\beta_1^{(t)}$ is its coefficient vector at iteration $t$, $\phi_j$ is the area-level design vector for cell $j$, given by the basis values for the area containing cell $j$, $\eta^{(t)}$ is the shared area-level effect, and $\tau_1^{(t)}$ scales the shared effect in the Gaussian component. Let $\sigma^{2(t)}$ denote the Gaussian residual variance at iteration $t$. Given $\theta_{1j}^{(t)}$ and $\sigma^{2(t)}$, we draw a single cell mean
\[
m_{kj}^{(t)} \sim \mathcal N\!\left(\theta_{1j}^{(t)},\,\frac{\sigma^{2(t)}}{N_{kj}}\right),
\]
and form the area-level draw

\[
\mu_{1k}^{(t)}
= \frac{\sum_{j\in\mathcal J_k} N_{kj}\, m_{kj}^{(t)}}{\sum_{j\in\mathcal J_k} N_{kj}}
= \sum_{j\in\mathcal J_k} q_{kj}\, m_{kj}^{(t)}.
\]
Repeating this over all iterations yields $\{\mu_{1k}^{(t)}\}_{t=1}^T$, and the posterior mean estimate is

\[
\widehat{\mu}_{1k}=\frac{1}{T}\sum_{t=1}^T \mu_{1k}^{(t)}.
\]
An algebraically equivalent approach draws $\mu_{1k}^{(t)}$ directly from a weighted Normal distribution; see the Appendix.

For the binomial outcome, let $p_j^{(t)}$ denote the cell-level success probability at iteration $t$
\[
p_j^{(t)}=\operatorname{logit}^{-1}\!\bigl(x_{2j}^\top \beta_2^{(t)}+\phi_j^\top \eta^{(t)} + \phi_j^\top \zeta^{(t)}\bigr),
\]
where $x_{2j}$ is the binomial covariate vector for cell $j$ and $\beta_2^{(t)}$ is its coefficient vector at iteration $t$. At each iteration, we then draw the cell-level total
\[
y_{kj}^{(t)} \sim \mathrm{Bin}\!\bigl(N_{kj},\,p_j^{(t)}\bigr).
\]
We aggregate these cell totals to obtain the area-level rate \[ \pi_k^{(t)}=\frac{\sum_{j\in\mathcal J_k} y_{kj}^{(t)}}{\sum_{j\in\mathcal J_k} N_{kj}}. \]
Collecting $\{\pi_k^{(t)}\}_{t=1}^T$ over iterations yields posterior draws for the area-level rate, and the posterior mean estimate is
\[
\widehat{\pi}_k=\frac{1}{T}\sum_{t=1}^T \pi_k^{(t)}.
\]
Credible intervals and interval scores are computed from the poststratified draws.

\section{Empirical Simulation Study}\label{sec: simulation}
To evaluate the performance of the proposed joint model, we carry out an empirical simulation that treats the public-use micro-data as the underlying population.  We begin with the 2023 one-year ACS PUMS data from Illinois. The raw extract contains 103,733 units.

\subsection{Simulation study design}
\label{sec:sim_design}
In the ACS PUMS data, we consider two unit-level outcomes: personal income in the past 12 months, measured in dollars (\texttt{PINCP}), as the Gaussian response, and poverty status as the binary response. Individuals with \texttt{PINCP}$=0$, for which the logarithmic transformation is undefined, and records with missing values in the analysis variables are removed. The resulting dataset contains $N = 92{,}817$ units and is treated as the finite population.

The Gaussian response is defined as the logarithm of \texttt{PINCP}. To place this response on the same numerical scale as the Bernoulli outcome and to improve numerical stability in the MCMC sampling, we apply a min–max transformation
\[
Z_{1i}
=
\frac{\log(\texttt{PINCP}_i) - \min\log(\texttt{PINCP})}
     {\max\log(\texttt{PINCP}) - \min\log(\texttt{PINCP})},
\]
so that $Z_{1i} \in [0,1]$ and roughly follows a Gaussian distribution.
The Bernoulli response indicates poverty status and is derived from the income-to-poverty ratio variable \texttt{POVPIP}. In the ACS, each individual is assigned an official poverty threshold by the Census Bureau, and poverty status is determined by comparing the individual’s income with this threshold. These thresholds vary by individual characteristics, such as age composition, are uniform across geographic areas, and are updated annually for inflation using the Consumer Price Index.

The variable \texttt{POVPIP} encodes this comparison as the ratio of an individual’s income to the applicable poverty threshold. We define the Bernoulli outcome as $Z_{2i}=1$ if $\texttt{POVPIP}_i < 100$, indicating income below the relevant threshold, and $Z_{2i}=0$ otherwise. As a result, poverty status is not determined by income alone and individuals with the same income may face different poverty thresholds and therefore different poverty classifications.

All predictors are categorical, consistent with the poststratification framework described previously.
Sex (\texttt{SEX}) is coded as $0$ for male and $1$ for female.
Educational attainment is derived from \texttt{SCHL} by collapsing categories into two groups: values below 21 correspond to education below a bachelor’s degree, and values of 21 or higher correspond to a bachelor’s degree or above.
After preprocessing, each unit is characterized by a bivariate response $(Z_{1i}, Z_{2i})$ and two categorical covariates.

From this finite population, we draw a fixed-size sample of size $n=1000$ using unequal-probability probability proportional to size (PPS) sampling without replacement, implemented via systematic PPS sampling. Let
\[
I_{\text{poor},i}=\mathbf{1}\{\texttt{POV}_i=1\},
\]
where $\texttt{POV}_i$ is the binary poverty indicator derived from $\texttt{POVPIP}_i$, and let $\texttt{PWGTP}_i$ denote the ACS person weight for unit i, that is, the number of population units represented by individual i. We define the PPS size measure as
$M_i = \texttt{PWGTP}_i \{1 + 5 I_{\text{poor},i}\}$. So that, holding $\texttt{PWGTP}_i$ fixed, units below the poverty threshold have six times the size measure of those above it. The first-order inclusion probabilities are then set to
\[
\pi_i \;:=\;
\frac{n\, M_i}{\sum_{k=1}^{N} M_k},
\qquad \text{so that } \sum_{i=1}^{N}\pi_i=n.
\]
Sampling is then conducted without replacement using systematic PPS sampling, which yields an exact sample size of $n=1000$ in each replicate. Because the inclusion probabilities depend on poverty status through the adjusted size measure $M_i$, the resulting design is informative.

Each sampled unit is assigned the design weight $w_i = 1/\pi_i$.
Following \citet{SavitskyToth2016}, we rescale these weights prior to model fitting
\[
\tilde w_i \;=\;
\frac{n\, w_i}{\sum_{j\in S} w_j}.
\]
This rescaling preserves relative differences across units while controlling the total information contributed by the pseudo-likelihood.

During poststratification (Section~\ref{sec:post}), known population cell counts $N_{kj}$ are used to aggregate unit-level predictions to the area level. The ground-truth in the simulation study are the finite-population PUMA-level mean of transformed log income and the finite-population PUMA-level poverty rate, computed from the full Illinois PUMS population after preprocessing.
This step shifts inference from the sampled units and their survey weights to the full population structure represented by census counts. For each model, we ran the Gibbs sampler for 2000 iterations, discarded the first 1000 as burn-in.
\subsection{Empirical Simulation Results}
\label{sec:sim_results}

In addition to mean–squared error (MSE), we evaluate uncertainty quantification using the 95\% interval score of \citet{gneiting2007strictly}.  For a central $(1-\alpha)$ posterior interval $[l,u]$ targeting a quantity $\theta$, the interval score is
\[
S^{\mathrm{int}}_\alpha(l,u;\theta)
=
(u - l)
+\frac{2}{\alpha}(l-\theta)\,\mathbf1\{\theta<l\}
+\frac{2}{\alpha}(\theta-u)\,\mathbf1\{\theta>u\},
\]
with $\alpha=0.05$.  Smaller values indicate intervals that are both shorter and well calibrated.  We report the sample–average score over 100 independent replicates for each of the 88 PUMAs.

Table~\ref{tab:sim_basis} summarize the empirical simulation results under the adjacency-based basis specifications. For each specification and response, area-specific MSE, interval score (IS), and empirical coverage rate (CR) are first computed over the 100 simulation replicates and then averaged across the 88 PUMAs. 

\begin{table}[htbp!]
\centering
\small
\caption{Empirical simulation results under the adjacency-based basis specification. Ratio to HT is the average of the area-specific MSE ratios relative to the Horvitz--Thompson estimator. Smaller values are better for MSE and interval score. Coverage rates closer to the nominal 95\% level indicate better calibration. MSE values are reported on the scale of $\times 10^{-3}$.}
\label{tab:sim_basis}
\begin{tabular}{llcccc}
\toprule
Response & Model & MSE & Ratio to HT & IS & CR \\
\midrule
Bernoulli & HT         & 7.834 & 1.000 & 0.422    & --    \\
Bernoulli & Univariate & 1.271 & 0.227 & 0.189 & 0.875 \\
Bernoulli & Multi-type & 1.197 & 0.220 & 0.158 & 0.919 \\
\addlinespace
Gaussian  & HT         & 1.517 & 1.000 & 0.190    & --    \\
Gaussian  & Univariate & 0.533 & 0.378 & 0.130 & 0.863 \\
Gaussian  & Multi-type & 0.212 & 0.154 & 0.073 & 0.908 \\
\bottomrule
\end{tabular}
\end{table}

Under the adjacency-based basis specification, the model-based estimators substantially outperform the Horvitz--Thompson estimator. For the Gaussian response, the multi-type model provides a clear improvement over the univariate model. The corresponding decreases are from $0.533\times 10^{-3}$ to $0.212\times 10^{-3}$ for MSE and from $0.130$ to $0.073$ for interval score. For the Bernoulli response, the average Bernoulli MSE decreases from $1.271\times 10^{-3}$ to $1.197\times 10^{-3}$, the average interval score decreases from $0.189$ to $0.158$, and the empirical coverage rate increases from $0.875$ to $0.919$. Taken together, these results show that the multi-type model improves overall performance compared with univariate model.


\section{ACS Microdata Analysis for Illinois}
\label{sec:data}
To evaluate the performance of our method under larger sample sizes, we illustrate our methodology using the 2023 ACS PUMS data for Illinois described below. The analysis targets 88 Public Use Microdata Areas (PUMAs) in Illinois, treating each as a small area or domain of interest. We removed all zero and missing income values from the dataset, resulting in a cleaned population dataset with 92,817 records. To maintain consistency with the poststratification steps described later, we exclusively selected units where the individual's age was 18 or older. This additional filtering criteria resulted in a final analysis dataset containing 91,208 records.

The outcomes and covariates are the same as in the simulation study, but the Illinois application is restricted to adults aged 18 and older to match the available poststratification tabulations. Our analysis focuses on two key outcomes as in the simulation study: (1) log-transformed individuals' income (treated as approximately Gaussian after transformation and min-max scaling), and (2) poverty status (coded as 1 if the income-to-poverty ratio \texttt{POVPIP} $<$ 100, and 0 otherwise). For poststratification and covariate adjustment, we use sex and education level (\texttt{SCHL}) as categorical predictors, ensuring all models employ the same set of covariates for consistency and interpretability.

\subsection{Model Implementation and Poststratification}

We fit the univariate and multi-type pseudo-likelihood models in Section~\ref{sec:method} using Gibbs sampling. At each MCMC iteration, cell-level posterior predictions are generated for all covariate strata within each PUMA and then aggregated to the PUMA level using the poststratification procedure in Section~\ref{sec:post}. This yields posterior draws for each PUMA-level estimand and carries model uncertainty from the cell level to the domain level.

To construct the poststratification weights, we use external population counts from the 2023 ACS 1-year PUMA-level tabulations rather than the PUMS microdata. The poststratification cells are defined by PUMA, sex, and bachelor’s degree status, restricted to individuals aged 18 and older to match the analysis population. These external counts provide the population totals $N_{kj}$ for all cells and ensure that the final PUMA-level estimates reflect the local demographic composition. In practice, most of the computational cost arises from sampling the P\'olya–Gamma latent variables. Because this step is also required in the univariate binomial model, the total runtime of the multi-type model is similar to that of the univariate binomial fit.

\subsection{Results and Model Comparison}

\begin{figure}[htbp!]
    \centering
    \includegraphics[width=\linewidth]{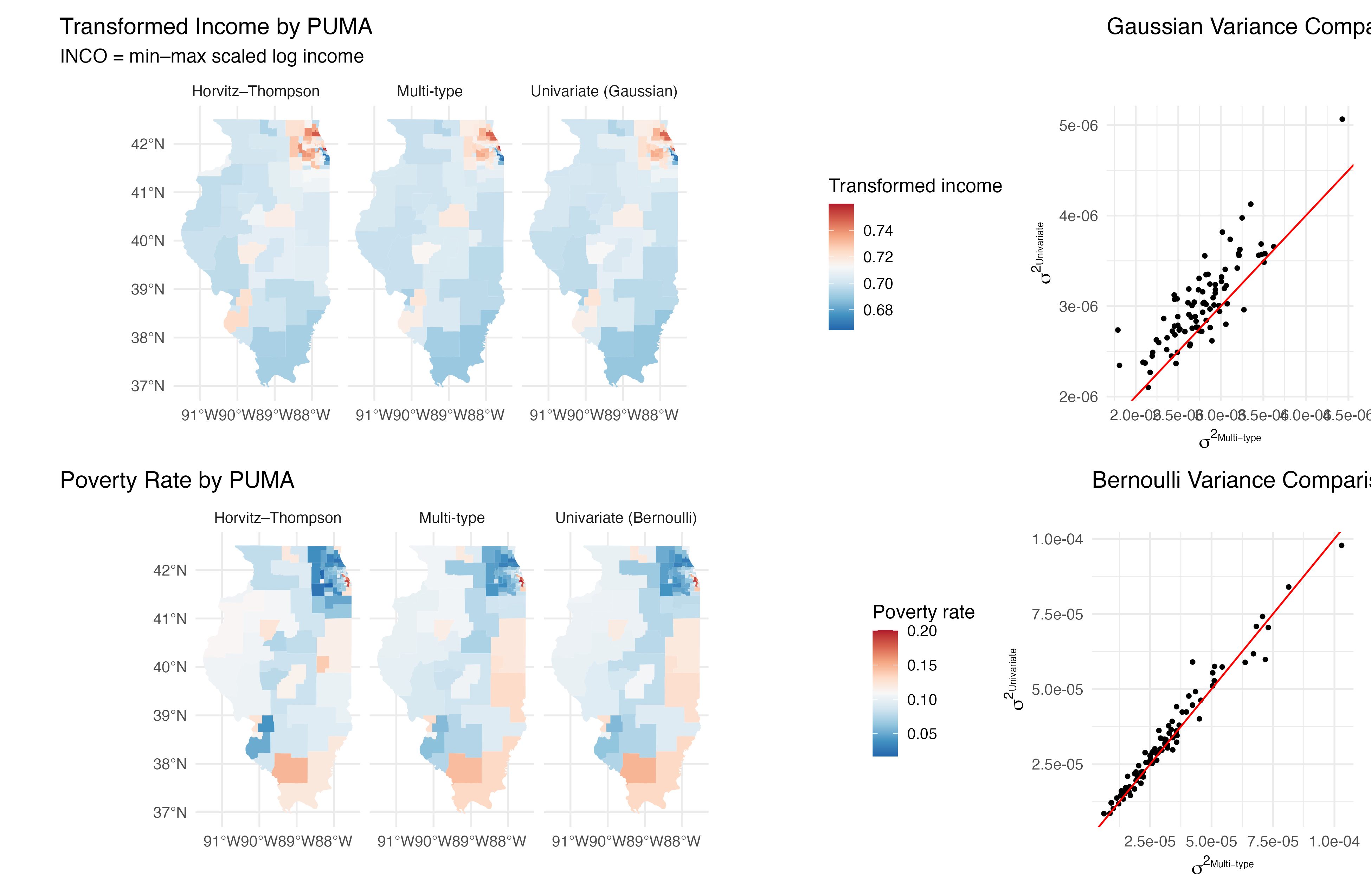}
    \caption{Illinois PUMA-level results under the adjacency-based basis specification. Top row: transformed income maps and Gaussian posterior variance comparison. Bottom row: poverty rate maps and Bernoulli posterior variance comparison. In each row, the first three panels show the Horvitz--Thompson estimator, the multi-type posterior mean, and the univariate posterior mean. The rightmost panel compares posterior variances from the multi-type and univariate models. Points above the 45-degree line indicate smaller posterior variance under the multi-type model. Here, INCO denotes min--max scaled log income.}
    \label{fig:region_basis_main}
\end{figure}

Figure~\ref{fig:region_basis_main} shows the Illinois PUMA-level results under the adjacency-based basis function specification. For both responses, the posterior mean maps from the multi-type and univariate models are close to the Horvitz--Thompson estimates. This indicates that all three methods produce similar mean surfaces and preserve the main spatial patterns in the data.
The main difference appears in posterior uncertainty. In both variance comparison panels, most points lie above the 45-degree line, indicating that the multi-type model generally yields smaller posterior variances than the univariate model. Specifically, the multi-type posterior variance is smaller than the univariate posterior variance in 84.1\% of PUMAs for the Gaussian response and in 72.7\% of PUMAs for the Bernoulli response. The reduction is more clear for the Gaussian response, but it is also present for the Bernoulli response. 
In our empirical simulation, the multi-type basis model achieved lower MSEs and interval scores. Building on those results, this application confirms that borrowing strength from a correlated response improves precision and reduces posterior uncertainty.

\section{Conclusion}\label{sec:conclusion}
We propose a Bayesian hierarchical model for jointly estimating unit-level Gaussian and binomial outcomes from microdata. The two outcomes are linked through a shared area-level random effect, allowing information to be borrowed across responses, while domain-specific heterogeneity is captured through spatial basis functions. The Gaussian outcome, defined as log-transformed income, is scaled to the $[0,1]$ range to improve mixing and to place it on a comparable scale to the binomial response. This approach extends unit-level small area estimation to settings with mixed data types, spatial dependence, and informative sampling. By combining pseudo-likelihood weighting for the complex survey design with P\'olya--Gamma augmentation for the binomial component, the model retains fully conjugate Gibbs updates and remains computationally efficient.

Area-level estimates are obtained through poststratification. In our empirical simulations, the multi-type basis model outperforms both the univariate specification and the Horvitz--Thompson estimator. For the Gaussian response, the joint model achieves lower MSE and smaller interval scores. For the Bernoulli response, the reduction in MSE is more moderate, but the interval score still shows clear reduction. In the Illinois application, the multi-type model produces mean surfaces similar to the direct estimates and univariate model while yielding smaller posterior variances for both outcomes. Together, these results show that borrowing strength across correlated responses can improve estimation precision and uncertainty quantification.

The basis specification provides a natural dimension-reduction strategy while preserving local spatial structure. Future work may consider other spatial representations, such as Moran eigenvectors or SPDE-type models, to further reduce computational burden when the number of areas is large. Overall, the proposed framework provides a practical approach to multi-type unit-level small area estimation under complex survey designs.

\section*{Acknowledgments}
  This article is released to inform interested parties of ongoing research and to encourage discussion. The views expressed on statistical issues are those of the authors and not those of the NSF. This research was partially supported by the U.S. National Science Foundation (NSF) under NSF grants NCSE-2215168 and NCSE-2215169. Jonathan Bradley's research was partially supported by NSF-DMS.

\bibliography{bghm}

@article{parker2022computationally,
  title={Computationally efficient {B}ayesian unit-level models for non-{G}aussian data under informative sampling with application to estimation of health insurance coverage},
  author={Parker, Paul A and Holan, Scott H and Janicki, Ryan},
  journal={The Annals of Applied Statistics},
  volume={16},
  number={2},
  pages={887--904},
  year={2022},
  publisher={Institute of Mathematical Statistics}
}

@article{bell16,
  title={An overview of the {US} {C}ensus {B}ureau's small area income and poverty estimates program},
  author={Bell, William R. and W. Basel, Wesley and J. Maples, Jerry},
  journal={Analysis of poverty data by small area estimation},
  pages={349--378},
  year={2016},
  publisher={Wiley Online Library}
}

@article{clinch2026exact,
  title={Exact Bayesian Inference for Multivariate Spatial Data of Any Size with Application to Air Pollution Monitoring: M. Clinch, JR Bradley},
  author={Clinch, Madelyn and Bradley, Jonathan R},
  journal={Journal of Agricultural, Biological and Environmental Statistics},
  pages={1--21},
  year={2026},
  publisher={Springer}
}

@article{bradley2022joint,
  title={Joint Bayesian analysis of multiple response-types using the hierarchical generalized transformation model},
  author={Bradley, Jonathan R},
  journal={Bayesian Analysis},
  volume={17},
  number={1},
  pages={127--164},
  year={2022},
  publisher={International Society for Bayesian Analysis}
}

@article{nandy2022bayesian,
  title={Bayesian Hierarchical Models for multi-type survey data using spatially correlated covariates measured with error},
  author={Nandy, Saikat and Holan, Scott H and Bradley, Jonathan R and Wikle, Christopher K},
  journal={arXiv preprint arXiv:2211.09797},
  year={2022}
}

@article{gneiting2007strictly,
  title={Strictly proper scoring rules, prediction, and estimation},
  author={Gneiting, Tilmann and Raftery, Adrian E},
  journal={Journal of the American statistical Association},
  volume={102},
  number={477},
  pages={359--378},
  year={2007},
  publisher={Taylor \& Francis}
}

@article{Argyriou,
	title = "Multi-task feature learning",
	journal = "Advances in neural information processing systems",
	year = "2007",
	volume="19",
	author = "A. Argyriou and T. Evgeniou and M. Pontil"
}

@article{Kim,
	title = "Statistical estimation of correlated genome associations to a quantitative trait network",
	journal = "PLos Genetics",
	year = "2009",
	volume="5",
	author = "S. Kim and E. P. Xing"
}

@article{Yang,
	title = "Heterogeneous multitask learning with joint sparsity constraints",
	journal = "NIPS",
	year = "2009",
	pages="2151–2159",
	author = "X. Yang and S. Kim and E. P. Xing"
}

@article{forest,
	title = "Stable graphical
	model estimation with random forests for discrete, continuous, and mixed variables",
	journal = "Computational
	Statistics amd Data Analysis",
	year = "2013",
	volume = "64",
	pages="132–152",
	author = "B. Fellinghauer and P. Buhlmann and M. Ryffel and M. Von Rhein and J. D. Reinhardt"
}

@article{copula2,
	title = "High-dimensional semiparametric
	gaussian copula graphical models",
	journal = "The Annals of Statistics",
	year = "2012",
	volume="40",
	pages="2293-2326",
	author = "H. Liu and F. Han and M. Yuan and J. Lafferty and L. Wasserman"
}

@article{rank1,
	title = "Regularized rank-based estimation of high-dimensional nonparanormal
	graphical models",
	journal = "The Annals of Statistics",
	year = "2012",
	volume="40",
	pages="2541-2571",
	author = "L. Xue and H. Zou"
}

@article{christensen2002,
	title = "Latent variable analysis of multivariate spatial data",
	journal = "Journal of the American Statistical Association",
	volume = "97",
	year = "2002",
	pages = "302-317",
	author = "W. F. Christensen and Y. Amemiya"
}

@article{gs2,
	title = "Multilevel latent Gaussian process model for mixed discrete and continuous multivariate response data",
	journal = "Journal of agricultural biological and environmental statistics",
	volume = "18",
	year = "2013",
	pages = "492-513",
	author = "E. M. Schliep and J. A. Hoeting"
}

@article{Todd,
	title = " A spatiotemporal multispecies model of a semicontinuous response",
	journal = "Journal of the Royal Statistical Society: Series C (Applied
	Statistics)",
	volume = "67",
	year = "2018",
	pages = "705-722",
	author = "C. M. Todd and B. Swallow and J. B. Illian and M. Toms"
}

@article{leClarke,
	title = "Generalized joint attribute modeling for biodiversity analysis: Median-zero, multivariate, multifarious data",
	journal = "Ecological Monographs",
	volume = "87",
	year = "2017",
	pages = "34-56",
	author = "J. S. Clarke and D. Nemergut and B. Seyednasrollah and P. Turner and S. Zhang"
}

@article{wu2015bayesian,
	title={Bayesian binomial mixture models for estimating abundance in ecological monitoring studies},
	author={Wu, Guohui and Holan, Scott H and Nilon, Charles H and Wikle, Christopher K and others},
	journal={The Annals of Applied Statistics},
	volume={9},
	number={1},
	pages={1--26},
	year={2015},
	publisher={Institute of Mathematical Statistics}
}

@article{copula1,
	title = "Copula Gaussian graphical models and their application to modeling functional disability data",
	journal = "The Annals of Statistics",
	year = "2011",
	volume="5",
	pages="969-993",
	author = "A. Dobra and A. Lenkoski"
}

@article{rank2,
	title = "The nonparanormal: Semiparametric estimation of high dimensional undirected graphs",
	journal = "The Journal of Machine Learning Research",
	year = "2009",
	volume="10",
	pages="2295-2328",
	author = "H. Liu and J. Lafferty and L. Wasserman"
}

@article{allen,
	title = " A general framework for mixed graphical models",
	journal = "arXiv:1411.0288",
	year = "2014",
	author = "E. Yang and P. Ravikumar and G. I. Allen and  Y. Baker and Y. -W. Wan and Z. Liu"
}

@article{polson2013bayesian,
  title={Bayesian inference for logistic models using P{\'o}lya--Gamma latent variables},
  author={Polson, Nicholas G and Scott, James G and Windle, Jesse},
  journal={Journal of the American statistical Association},
  volume={108},
  number={504},
  pages={1339--1349},
  year={2013},
  publisher={Taylor \& Francis}
}

@article{ekvall2022mixed,
  title={Mixed-type multivariate response regression with covariance estimation},
  author={Ekvall, Karl Oskar and Molstad, Aaron J},
  journal={Statistics in Medicine},
  volume={41},
  number={15},
  pages={2768--2785},
  year={2022},
  publisher={Wiley Online Library}
}

@article{MolinaRao2010,
  author  = {Molina, Isabel and Rao, J. N. K.},
  title   = {Small Area Estimation of Poverty Indicators},
  journal = {The Canadian Journal of Statistics},
  year    = {2010},
  volume  = {38},
  number  = {3},
  pages   = {369--385},
  doi     = {10.1002/cjs.10051}
}

@article{ParkerJanickiHolan2023,
  author  = {Parker, Paul A. and Janicki, Ryan and Holan, Scott H.},
  title   = {A Comprehensive Overview of Unit-Level Modeling of Survey Data for Small Area Estimation Under Informative Sampling},
  journal = {Journal of Survey Statistics and Methodology},
  year    = {2023},
  volume  = {11},
  number  = {4},
  pages   = {829--857},
  doi     = {10.1093/jssam/smad020}
}

@article{SavitskyToth2016,
  author  = {Savitsky, Terrance D. and Toth, Daniell},
  title   = {Bayesian Estimation Under Informative Sampling},
  journal = {Electronic Journal of Statistics},
  year    = {2016},
  volume  = {10},
  number  = {1},
  pages   = {1677--1708},
  doi     = {10.1214/16-EJS1153}
}

@article{hidiroglou2016comparison,
  title={Comparison of unit-level and area-level small area estimators},
  author={Hidiroglou, Michael A and You, Yong},
  journal={Survey Methodology},
  volume={42},
  number={1},
  pages={41--61},
  year={2016}
}

@article{parker2020conjugate,
  title={Conjugate Bayesian unit-level modelling of count data under informative sampling designs},
  author={Parker, Paul A and Holan, Scott H and Janicki, Ryan},
  journal={Stat},
  volume={9},
  number={1},
  pages={e267},
  year={2020},
  publisher={Wiley Online Library}
}

@article{bugallo2024small,
  title={Small area estimation of labour force indicators under unit-level multinomial mixed models},
  author={Bugallo, M and Morales, D and Esteban, MD and P{\'e}rez-Mart{\'\i}n, A and Hobza, T},
  journal={Journal of the Royal Statistical Society Series A: Statistics in Society},
  volume={187},
  number={1},
  pages={241--269},
  year={2024},
  publisher={Oxford University Press}
}

@article{bradley2015multivariate,
  title={Multivariate spatio-temporal models for high-dimensional areal data with application to longitudinal employer-household dynamics},
  author={Bradley, Jonathan R and Holan, Scott H and Wikle, Christopher K},
  year={2015}
}

@article{bradley2016multivariate,
  title={Multivariate spatio-temporal survey fusion with application to the American Community Survey and Local Area Unemployment Statistics},
  author={Bradley, Jonathan R and Holan, Scott H and Wikle, Christopher K},
  journal={Stat},
  volume={5},
  number={1},
  pages={224--233},
  year={2016},
  publisher={Wiley Online Library}
}

@article{hobza2018small,
  title={Small area estimation of poverty proportions under unit-level temporal binomial-logit mixed models},
  author={Hobza, Tom{\'a}{\v{s}} and Morales, Domingo and Santamar{\'\i}a, Laureano},
  journal={Test},
  volume={27},
  number={2},
  pages={270--294},
  year={2018},
  publisher={Springer}
}

@article{you2002pseudo,
  title={A pseudo-empirical best linear unbiased prediction approach to small area estimation using survey weights},
  author={You, Yong and Rao, JNK1944372},
  journal={Canadian Journal of Statistics},
  volume={30},
  number={3},
  pages={431--439},
  year={2002},
  publisher={Wiley Online Library}
}

@article{porter2015small,
  title={Small area estimation via multivariate Fay--Herriot models with latent spatial dependence},
  author={Porter, Aaron T and Wikle, Christopher K and Holan, Scott H},
  journal={Australian \& New Zealand Journal of Statistics},
  volume={57},
  number={1},
  pages={15--29},
  year={2015},
  publisher={Wiley Online Library}
}
\bibliographystyle{apalike}
\appendix

\newpage
\section*{Appendix}\label{app}

\subsection*{Full Conditional Distribution for the Shared Random Effect}

We derive the posterior distribution for $\eta$ in a step-by-step manner. Let
\[
Z_{1,i} = x_{1,i}^\top \beta_1 + \tau_1 \boldsymbol{\phi}_i^\top \eta + \varepsilon_i,
\,\,\,\,
\varepsilon_i \sim N(0,\sigma^2),
\]
and
\[
\mathrm{logit}(p_i)=\psi_i=x_{2,i}^\top \beta_2 + \boldsymbol{\phi}_i^\top \eta + \boldsymbol{\phi}_i^\top \zeta,
\,\,\,\,
Z_{2,i}\mid p_i \sim \mathrm{Binomial}(n_i,p_i).
\]
We introduce P\'olya--Gamma latent variables
\[
\omega_i \sim \mathrm{PG}(b_i,\psi_i),\,\,\,\, b_i=\tilde w_i n_i,
\]
and define
\[
\kappa_i=\tilde w_i\left(Z_{2,i}-\frac{n_i}{2}\right).
\]
Let
\[
\Phi=(\boldsymbol{\phi}_i^\top)_{i\le n}\in\mathbb{R}^{n\times q},\,\,\,\,
W=\mathrm{diag}(\tilde w_1,\ldots,\tilde w_n),\,\,\,\,
\Omega=\mathrm{diag}(\omega_1,\ldots,\omega_n).
\]
Further define
\[
a_G = Z_1 - X_1\beta_1,\,\,\,\,
a_B = X_2\beta_2 + \Phi\zeta,
\]
and assume the prior
\[
\eta \sim N(0,\sigma_\eta^2 I_q).
\]
The Gaussian pseudo-likelihood contributes
\[
\ell_G(\eta)\doteq
-\frac{1}{2}\eta^\top\left(\frac{\tau_1^2}{\sigma^2}\Phi^\top W \Phi\right)\eta
+\eta^\top\left(\frac{\tau_1}{\sigma^2}\Phi^\top W a_G\right).
\]
The augmented likelihood contributes
\[
\ell_B(\eta)\doteq
-\frac{1}{2}\eta^\top(\Phi^\top\Omega \Phi)\eta
+\eta^\top\left(\Phi^\top\kappa - \Phi^\top\Omega a_B\right).
\]
The Gaussian prior for $\eta$ gives
\[
\log p(\eta)\doteq -\frac{1}{2}\eta^\top\left(\frac{1}{\sigma_\eta^2}I_q\right)\eta.
\]
Adding the contributions, we obtain
\[
\log\pi(\eta\mid \cdot)\doteq -\frac{1}{2}\eta^\top Q\eta + \eta^\top h,
\]
where
\[
Q=\frac{\tau_1^2}{\sigma^2}\Phi^\top W \Phi + \Phi^\top\Omega \Phi + \frac{1}{\sigma_\eta^2}I_q,
\,\,\,\,
h=\frac{\tau_1}{\sigma^2}\Phi^\top W a_G + \Phi^\top\kappa - \Phi^\top\Omega a_B.
\]
Completing the square shows that
\[
\eta\mid \cdot\sim N(\mu_\eta,\Sigma_\eta),\,\,\,\,
\Sigma_\eta = Q^{-1},\,\,\,\,
\mu_\eta = \Sigma_\eta q.
\]
With the shorthand
\[
Q_G=\Phi^\top (W/\sigma^2) \Phi,\,\,\,\,
Q_B=\Phi^\top\Omega \Phi,\,\,\,\,
b_G=\Phi^\top W(Z_1-X_1\beta_1),
\]
\[
O_\kappa=\Phi^\top\kappa,\,\,\,\,
O_\zeta=\Phi^\top\Omega(X_2\beta_2+\Phi\zeta),
\]
the posterior can be written as
\[
\Sigma_\eta^{-1}=\tau_1^2 Q_G + Q_B + \frac{1}{\sigma_\eta^2}I_q,
\,\,\,\,
\mu_\eta=\Sigma_\eta\left(\frac{\tau_1}{\sigma^2}b_G + (O_\kappa - O_\zeta)\right).
\]

\subsection*{Equivalent method for the Gaussian outcome}
For completeness, we record an equivalent Normal formulation for the Gaussian poststratification step. Conditional on parameters at iteration $t$, the cell-level draws
\[
m^{(t)}_j \mid \theta^{(t)}_{1j}, \sigma^{2(t)} \sim
N\!\left(\theta^{(t)}_{1j}, \frac{\sigma^{2(t)}}{N_{kj}}\right),\,\,\,\, j\in J_k,
\]
are conditionally independent, where under the final model
\[
\theta^{(t)}_{1j}=x_{1j}^\top\beta^{(t)}_1+\tau^{(t)}_1\,\boldsymbol{\phi}_j^\top\eta^{(t)}.
\]
Because the area mean is a weighted sum,
\[
\mu^{(t)}_{1k}=\sum_{j\in J_k} q_j m^{(t)}_j,\,\,\,\,
q_j=\frac{N_{kj}}{\sum_{j\in J_k}N_{kj}},
\]
linearity of the Normal distribution implies
\[
\mu^{(t)}_{1k}\sim
N\!\left(\sum_{j\in J_k} q_j\theta^{(t)}_{1j},\ \sum_{j\in J_k} q_j^2\frac{\sigma^{2(t)}}{N_{kj}}\right)
=
N\!\left(\frac{\sum_{j\in J_k} N_{kj}\theta^{(t)}_{1j}}{\sum_{j\in J_k} N_{kj}},\
\frac{\sigma^{2(t)}}{\sum_{j\in J_k} N_{kj}}\right).
\]
Thus, rather than simulating each $m^{(t)}_j$ and averaging, one may draw $\mu^{(t)}_{1k}$
directly from this aggregate Normal distribution. The aggregate--then--generate and
generate--then--aggregate procedures are exactly equivalent.
\end{document}